\documentclass
[reprint,showpacs,preprintnumbers,amsmath,amssymb,aps,prl,twocolumn, superscriptaddress
]{revtex4-1}

\usepackage{color}
\usepackage{graphicx}
\usepackage{dcolumn}
\usepackage{bm}
\usepackage{amsmath}

\begin{document}

\bibliographystyle{apsrev} 

\title{Efficient generation of many-body entangled states by multilevel oscillations}

\author{Peng Xu}
\affiliation{School of Physics and Technology, Wuhan University, Wuhan, Hubei 430072, China}

\author{Su Yi}
\affiliation{CAS Key Laboratory of Theoretical Physics, Institute of Theoretical Physics, Chinese Academy of Sciences, P.O. Box 2735, Beijing 100190, China}
\affiliation{School of Physical Sciences \& CAS Center for Excellence in Topological Quantum Computation, University of Chinese Academy of Sciences, Beijing 100049, China}

\author{Wenxian Zhang}
\email[Corresponding email: ]{wxzhang@whu.edu.cn}
\affiliation{School of Physics and Technology, Wuhan University, Wuhan, Hubei 430072, China}

\date{\today}

\begin{abstract}
{We generate high-fidelity massively entangled states in an antiferromagnetic spin-1 Bose-Einstein condensate (BEC) by utilizing multilevel oscillations.} Combining the multilevel oscillations with additional adiabatic drives, we greatly shorten the necessary evolution time and relax the requirement on the control accuracy of quadratic Zeeman splitting, from micro-Gauss to milli-Gauss, for a $^{23}$Na spinor BEC. The achieved high fidelities over $96\%$ show that two kinds of massively entangled states, the many-body singlet state and the twin-Fock state, are almost perfectly generated. The generalized spin squeezing parameter drops to a value far below the standard quantum limit even with the presence of atom number fluctuations and stray magnetic fields, illustrating the robustness of our protocol under real experimental conditions. The generated many-body entangled states can be employed to achieve the Heisenberg-limit quantum precision measurement and to attack nonclassical problems in quantum information science.
\end{abstract}

\maketitle

Massive entanglement is of great importance for applications in quantum computing (e.g., logical qubit design utilizing decoherence-free subspace)~\cite{Haffner2008Quantum, Horodecki2009Quantum, Lidar1998Decoherence, West2010High}, quantum information processing~\cite{Sorensen2001Many, Cabello2002Particle, Prevedel2007Experimental, Pezze2009Entanglement}, and quantum metrology beyond the standard quantum limit~\cite{Wineland1992Spin, Wineland1994Squeezed, Gross2010nonlinear, Riedel2010atom, Wu2016Using, Feldmann2018Interferometric, Pezze2017Heralded, Smerzi2018Quantum}. For these applications, it is desirable to involve as many particles as possible into entangled states. Two well-known massively entangled states are many-body singlet state and twin-Fock state. For the many-body singlet state, in which a large number of nonzero spins consist of a ``giant" zero total spin, it has attracted a great amount of attention to enhance the sensitivity of a gradient magnetometer~\cite{Toth2013Macroscopic} and to realize robust logical qubits in decoherence-free subspace~\cite{Lidar1998Decoherence, West2010High, Prevedel2007Experimental}. For the twin-Fock state, with half of particles each in two orthogonal modes, it is often employed to improve the precision of a quantum magnetometer to the Heisenberg limit~\cite{Duan2013Generation, Luo2017Deterministic, L2011Twin, Kruse2016Improvement}.

However, these entangled states are typically very fragile. To generate these states in current experiments, the main challenge comes from the extremely fine control of the experimental conditions and deep suppression of the environmental noises~\cite{Luo2017Deterministic, Sun2017Efficient, Koashi2000Exact, Ho2000Fragmented}. For a $^{23}$Na antiferromagnetic spinor condensate, both the bias field and the stray magnetic fields in a laboratory must be below micro-Gauss in order to observe {its ground state} for $N\sim 1,000$~\cite{Koashi2000Exact, Ho2000Fragmented, Mueller2006Fragmentation, Jiang2014Mapping}. {As mentioned in previous papers, the antiferromagnetic spin-1 BEC exhibits two quantum phases}~\cite{Mueller2006Fragmentation, Liu2009Quantum, Jiang2014Mapping, Jacob2012Phase, Duan2018Classification, Bookjans2011Quantum, Vinit2017Precise, Frapolli2017Stepwise}. Ideally, by adiabatically tuning the quadratic Zeeman splitting from positive infinity through zero to negative infinity~\cite{Zhao2014Dynamics, Gerbier2006Resonant, Leslie2009Amplification}, one can respectively generate the many-body singlet state and the twin-Fock state by passing through critical point of quantum phase transition. The adiabaticity usually requires a finite and moderate energy gap between the ground and the first excited states. However, such a requirement is impossible to meet in the antiferromagnetic spin-1 BEC, because the gap reduces inversely proportional to the number of atoms $N$, $\Delta E \sim 1/N$~\cite{Bookjans2011Quantum, Jacob2012Phase, Duan2018Classification, Luigi2013Spin, Sala2016Shortcut, Zhao2018Lattice}, {{which drops faster than that in a ferromagnetic spin-1 BEC with $\Delta E \sim 1/N^{1/3}$~\cite{Duan2013Generation, Hoang2016Adiabatic, Luo2017Deterministic, Zou2018Beating, Xue2018Universal}}.} Indeed, given a $^{23}$Na BEC with $N=1,000$ atoms and a typical density of $10^{14}$ cm$^{-3}$ ($c'_2 \approx$ 25 Hz), the adiabatic evolution time to reach the ground state must be much larger than $N^3/(108c'_2)\sim 10^5$ seconds by a crude estimation, which is many orders of magnitude larger than the condensate lifetime of $\sim 100$ seconds~\cite{Koashi2000Exact, Sala2016Shortcut}. For almost two decades {since the prediction of the many-body singlet state by Law {\it et al} in 1998~\cite{Law1998Quantum}}, a practical and experimentally feasible method has been longed to generate this highly entangled state in an antiferromagnetic spinor BEC~\cite{Koashi2000Exact, Ho2000Fragmented, Kawaguchi2012Spinor, Stamper2013Spinor, Sala2016Shortcut, Sun2017Efficient, Zhao2018Lattice}.

\begin{figure}
\includegraphics[width=3.3in]{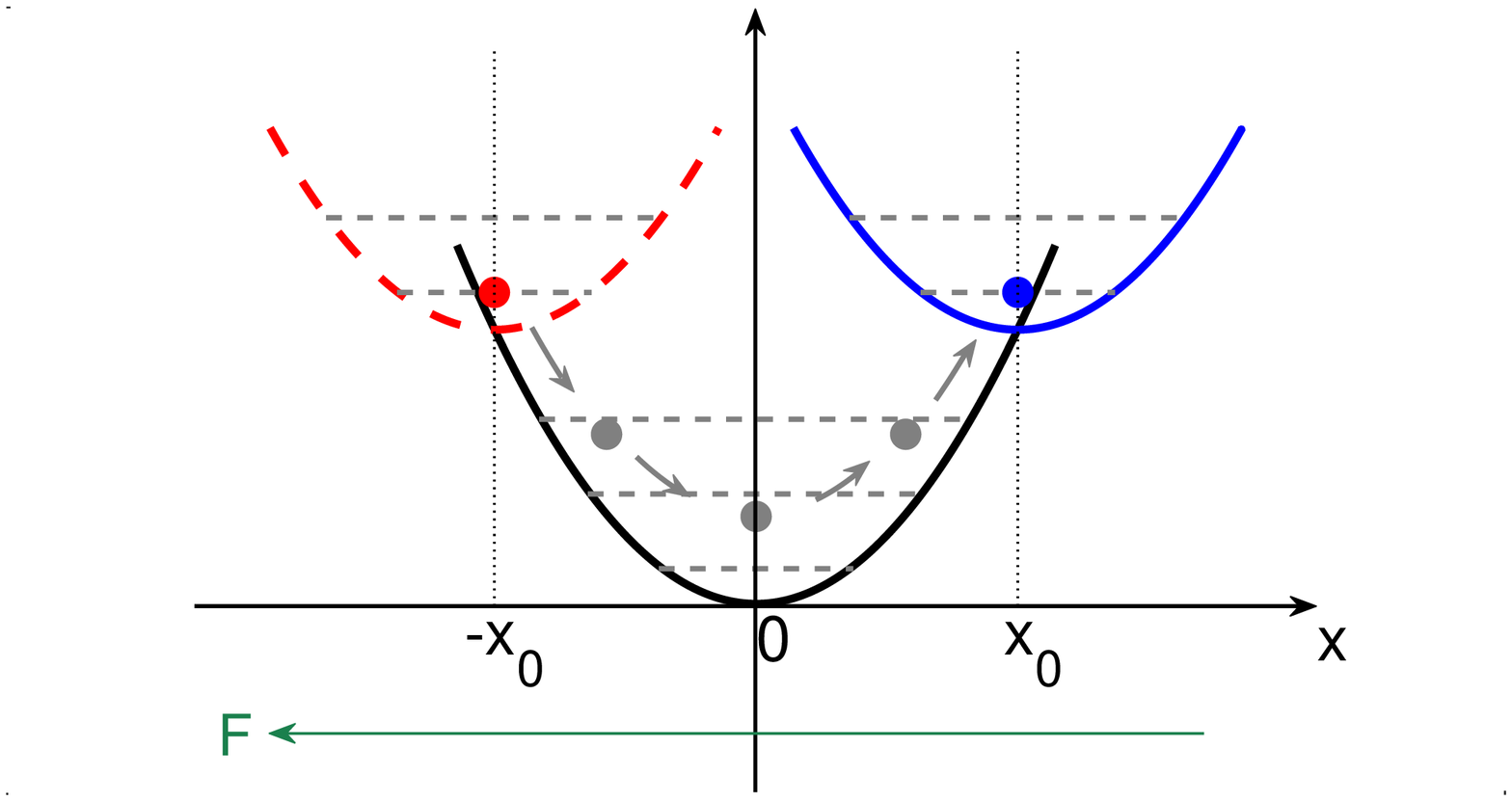}
\caption{\label{fig:1} (Color online.) Schematic of multilevel oscillation for a harmonic oscillator. An initial state (red circle) oscillates from $-x_0$ to $x_0$ in the harmonic potential (black line), and reaches the target state (blue circle). In the eigenenergy basis of the final harmonic potential (blue line), the initial wide probability distribution in many levels shrinks to a Kronecker $\delta$ distribution in a single (ground) level after the multilevel oscillation. The grey dashed lines denote the corresponding energy levels.}
\end{figure}

{In this Letter, we theoretically achieve the generation of massively entangled states, the singlet and twin-Fock states, in an antiferromagnetic $^{23}$Na spin-1 condensate by employing a rapid, efficient and robust method.} This method accelerates the dynamics and relaxes the requirement on the control accuracy of quadratic Zeeman splitting by partially replacing the adiabatic evolution near the quantum critical point of the phase transition with multilevel oscillations~\cite{Pu1999Spin, Zhang2005Coherent, Chang2007Number, Li2015Localizing, Bruce2011Manipulating, Claudon2004Coherent, Claudon2008Rabi}. We term the method as adiabatic and multilevel-oscillation (AMO) process for the generation of the singlet states, and as AMO and adiabatic (AMOA) process for the generation of the twin-Fock states.

The main advantage of the multilevel oscillation over an adiabatic process can be in principle illustrated by a harmonic oscillator as shown in Fig.~\ref{fig:1}. Consider an oscillator with a mass $M$ in an extra linear potential~\cite{Scully1997Quantum}
\begin{eqnarray}
\label{eq:so1}
H_o &=& \frac{P^2}{2M} + \frac{1}{2}M\omega^2 x^2 + V(x),
\end{eqnarray}
where $P$ is the momentum, $x$ the position, $\omega$ the trapping angular frequency, and $V(x) = Fx$ with $F$ an additional force applied on the oscillator. To reach the desired target state, one may employ an adiabatic process by slowly tilting the linear potential from $F_0x$ to $-F_0x$, or by a multilevel oscillation process by setting $F=0$ for a half period and then setting $F=-F_0$, as illustrated in Fig.~\ref{fig:1}.

It is easy to calculate the required adiabatic evolution time $T_A \gg 10\sqrt{2}/\omega$ if we set $x_0=10\sqrt{\hbar/(M\omega)}$ and the multilevel oscillation time $T_{MO} = \pi/\omega$. Clearly, the multilevel oscillation time is much shorter than the adiabatic one when the oscillator is transferred from $x=-x_0$ to $x_0$, thus the process is greatly accelerated. {In fact, the multilevel oscillation process is a generalized ``Rabi" oscillation for a half period in a multilevel system~\cite{note1}.}

\begin{figure}
\includegraphics[width=3.3in]{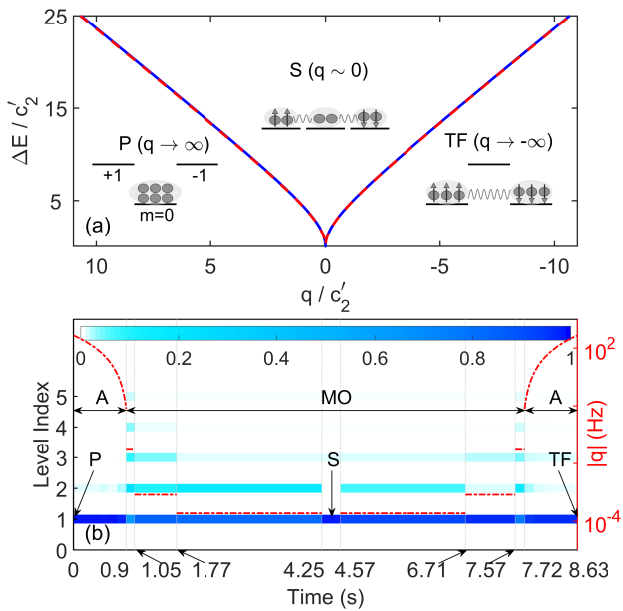}
\caption{\label{fig:2} (Color online.) (a) Phase diagram of an antiferromagnetic spin-1 BEC. Three critical quantum ground states (i.e. polar, singlet, and twin-Fock states) are illustrated by their atom distribution in three spin components. The blue solid line (red dashed line) denotes the energy gap $\Delta E$ for a total atom number $N=1,000$ ($N=100$). (b) AMO and AMOA processes in the instantaneous eigenenergy basis with A representing adiabatic and MO the multilevel oscillation. For the AMO process, starting from an initial polar state (P), the condensate evolves adiabatically at first, and after three multilevel oscillations, the system reaches the singlet state (S). For the AMOA process, three more multilevel oscillations are followed by another adiabatic evolution in order to generate the twin-Fock state (TF). The blue-and-cyan ribbons show the fidelities of the state on the instantaneous low-energy eigenstates for $N=1,000$. The color scales the fidelity. The red dot-dashed line describes $q$ (right axes) as a piecewise function of time. When $t<4.25$ s ($t>4.57$ s), $q>0$ ($q<0$); $q=0$ if $t \in [4.25, 4.57]$ s.}
\end{figure}

For an antiferromagnetic $^{23}$Na spin-1 condensate, the effective Hamiltonian under the single spatial mode approximation, {which is valid up to $10,000$ atoms,} is ($\hbar=1$)~\cite{Stamper-Kurn1998Optical, Stenger1998Spin, Ho1998Spinor, Law1998Quantum, Ohmi1998Bose, Pu1999Spin, Yi2002Single, Duan2013Generation, Luo2017Deterministic, Zou2018Beating, note1}
\begin{eqnarray}
\label{eq:1}
H_e=c'_2 \frac{\bm{L}^2}{N} -q a_0^{\dagger}a_0.
\end{eqnarray}
The first term describes the spin-exchange collision, where we set $c'_2=25$ Hz for a typical condensate density, and $\bm{L}\equiv \sum_{mn}{a}_{m}^{\dag}\bm{f}_{mn}{a}_{n}$ with $\bm{f}_{mn}$ the spin-1 matrices and $a_m (a_m^\dagger)$ the annihilation (creation) operator in spin component $m$. The second term represents the magnetic energy with $q$ the quadratic Zeeman splitting of a single atom. Depending on $q/c'_2$, the system exhibits two phases, resulting from the competition between the quadratic Zeeman term and the spin-exchange collision. Near the critical point where $q$ is small, the energy gap can be calculated perturbatively $\Delta E/c'_2 \simeq 6/N-0.1907\times N\times q+0.0253\times N^3\times q^2$~\cite{Luigi2013Spin}. The minimal gap occurs at the critical point $q_c/c'_2=3.7688/N^2$, very close to zero if $N$ is large, as shown in Fig.~\ref{fig:2}(a).

{The spin-1 BEC and the harmonic oscillator share the same chain-form Schr\"odinger equation except for different coefficients, as derived in the Supplemental Materail~\cite{note1, Bruce2011Manipulating, Luigi2013Spin}. However, the effective potential for the spin-1 BEC is anharmonic so that a single large-amplitude oscillation may take an infinitely long time~\cite{Zhang2005Coherent, Chang2007Number, Li2015Localizing, note1}. Instead, the total evolution time may be shorter if we stepwise change the quadratic Zeeman splitting $q$ so that the system evolves through many local harmonic oscillations.}

{Following the above strategy, we successfully generate with a high fidelity the many-body singlet state at $q=0$ and the twin-Fock state as $q\rightarrow -\infty$ by employing the AMO and AMOA processes respectively. The initial state is a polar state of a $^{23}$Na condensate, where all atoms are in the $m=0$ spin component of the $S=1$ ground hyperfine manifold. This polar state is easily accessible in experiment by setting a large bias magnetic field and optically pumping away the atoms in $m=\pm 1$ spin components~\cite{Jiang2016First, Luo2017Deterministic}. For a large but finite $q_0=277$ Hz, the initially prepared polar state overlaps with the ground state with a high fidelity about $1-(c'_2/q_0)^2/2$, which is over 99\%.}

The adiabatic process of the AMO is carried out numerically by slowly reducing $q$ according to
$
q(t)=q_0\times (1-{t}/{T_0})^2,
$
where $T_0=0.955$ s, and $t$ ends up at $0.9$ s with a final $q_f=0.788$ Hz. For convenience in experimental implementation, we linearly sweep the magnetic bias field thus a parabolic function for $q(t)$. In this adiabatic process as shown in Fig.~\ref{fig:2}(b), we calculate the adiabatic parameter $\beta=|\partial \langle e|q a^\dag_0a_0|g \rangle/\partial t|/\Delta E^2$, with the instantaneous ground state $|g\rangle$ and the first excited state $|e\rangle$ of the Hamiltonian in Eq.~(\ref{eq:1}). We find $\beta \le 0.054$ during the whole adiabatic process, thus the adiabatic condition is satisfactorily fulfilled since $\beta \ll 1$.

\begin{figure}
\includegraphics[width=3.3in]{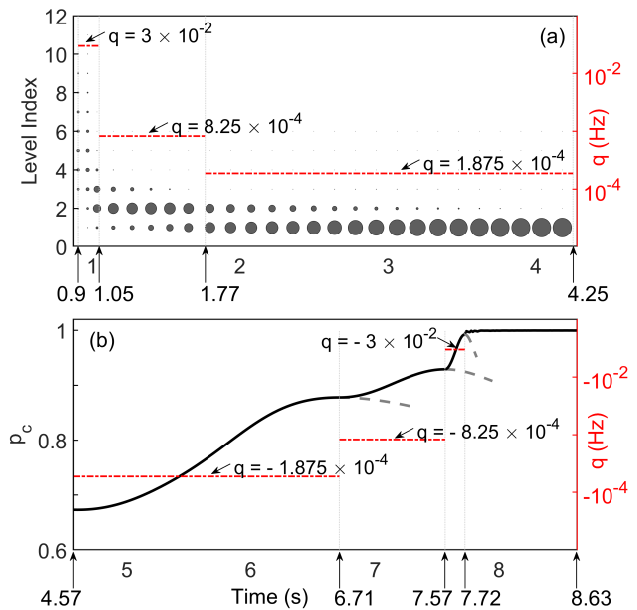}
\caption{\label{fig:3} (Color online.) (a) Dynamics of the occupied levels during the three multilevel oscillations for $N=1,000$, in the eigenenergy basis for $q=0$. The number of occupied levels decreases as time goes by ($q$ decreases piecewise). The radii of circles scale the fidelity. (b) Evolution of the conversion efficiency $p_c = (N-N_0)/N$ during the generation of the twin-Fock state for $N=1,000$. The process is divided into three multilevel oscillations and an additional adiabatic process, separated by three vertical grey dotted lines where the conversion efficiencies reach a local maximum. In both (a) and (b), the red dot-dashed lines represent the quadratic Zeeman splitting $q$ (right axis).}
\end{figure}

For the three multilevel oscillations as shown in Fig.~\ref{fig:2}(b) to generate the singlet state, we observe significant excitations in the instantaneous eigenenergy basis, indicating these multilevel oscillations are diabatic. To better understand this process, we redraw the probability distribution in the eigenenergy basis $|l\rangle$ for $q=0$ in Fig.~\ref{fig:3}(a). The optimized values of $q(t)$ and the corresponding evolution times are also listed~\cite{note1}. {Briefly, any state $|\psi\rangle$ is expanded as $\sum_{l=0}^N \beta_l |l\rangle$, and we define an eigenenergy level as occupied if $|\beta_l|^2>0.1\%$. For a state at a given time, we calculate $|\beta_l|^2>0.1\%$ and count the number of occupied levels $K$. The goal of stepwise multilevel oscillations is to reduce $K$ to 1, i.e., to the singlet state. In each multilevel oscillation, for a given constant $q$, we evolve the system and monitor $K$ till $K$ reaches its first local minimum $K(q)$; then we sweep $q$ to further minimize $K(q)$ in order to find the optimal range of $q$, as detailed in the SM~\cite{note1}.}

As shown in Fig.~\ref{fig:3}(a), $K$ shrinks from 15 to 4 during the first multilevel oscillation. The number further shrinks to 2 and 1, respectively, during the second and the third multilevel oscillations. Eventually, the fidelity of the final state (with respect to the singlet state) is over 99\%. We note that the required smallest magnetic field is about 0.8 mG, corresponding to $q=1.875\times 10^{-4}$ Hz. This field strength is easily accessible in experiments and about three orders of magnitude stronger than previous estimations of microGauss~\cite{Koashi2000Exact, Ho2000Fragmented}. Remarkably, the total evolution time is only 4.25 s, at least five orders of magnitude shorter than a full adiabatic process~\cite{Sala2016Shortcut}. {Here we show only one set of $q(t)$, while there are many other sets resulting in fast generating the singlet state with similar or even higher fidelity~\cite{note1}.}

After generating the singlet state, we employ a reversed procedure but with negative $q$ to produce the twin-Fock state, as shown in Figs.~\ref{fig:2}(b) and~\ref{fig:3}(b). We notice in Fig.~\ref{fig:2}(a) that the energy gap is almost symmetric about $q=0$, reminding us that the twin-Fock state may be reached by simply reverse the AMO process with only sign change of $q(t)$. This whole process is the AMOA process. Indeed, the evolved final state overlaps with the twin-Fock state with a fidelity higher than 96\%, indicating the success of the AMOA method. A so high efficiency contrasts sharply to a direct Landau-Zener transition by linearly sweeping $q$ from $q_0$ to $-q_0$ in the same time period 8.63 s, where the fidelity of the twin-Fock state is almost zero~\cite{Zener1998Non, Wittig2005Landau}.

As elegant as the above AMO and AMOA processes to efficiently generate the many-body singlet state and the twin-Fock state, a practical final state is never a pure one in a real experiment. To evaluate the robustness of the AMO process under realistic experimental conditions, we need to include the effects of the stray magnetic fields (both dephasing and relaxation effects), the atom number shot noise, and the atom loss during the evolution. {Although the control errors in $q$ and timing are non-negligible noise source, it is easy to prove that they are equivalent to the dephasing effect.} Furthermore, in a real experiment, it is almost impossible to measure the quantum state fidelity, thus we choose the generalized spin-squeezing parameter to monitor the AMO process. In addition, this parameter can also estimate the entanglement degree of the evolved quantum state. The parameter is defined as
\begin{eqnarray}
\label{eq:3}
\xi^2=\frac{1}{SN}\sum_{\alpha=x,y,z}(\Delta L_\alpha)^2,
\end{eqnarray}
where $(\Delta L_\alpha)^2 = \langle L_\alpha^2 \rangle - \langle L_\alpha \rangle^2$. A spin state is squeezed if $\xi^2<1$, compared to a coherent spin state with $\xi^2=1$ which sets the standard quantum limit. For the singlet state, $\xi^2=0$.

\begin{figure}
\includegraphics[width=3.3in]{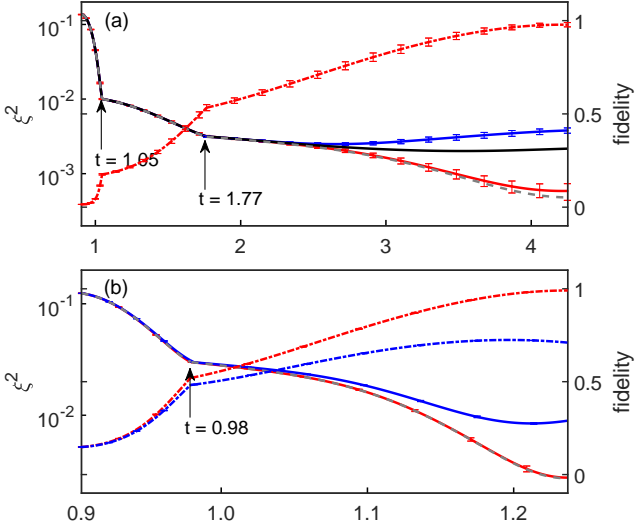}
\caption{\label{fig:4} (Color online.) {Dynamics of the generalized spin-squeezing parameter $\xi^2$ and fidelities of the state on the singlet state during the multilevel oscillation processes, (a) with fluctuations in the bias field and in the initial atom number for an average $\overline{N}=1,000$ where red (blue) solid line represents $\xi^2$ for even (odd) atom numbers and black solid line for the average, and the red dot-dashed line represents the fidelity for even $N$; (b) with relaxation and dephasing noises where the red solid (dot-dashed) line is for $\xi^2$ (fidelity), and with atom loss and dephasing noises where blue solid (dot-dashed) line is for $\xi^2$ (fidelity) for $N=100$.} The grey dashed lines represent the perfect multilevel oscillations without any noise.}
\end{figure}

First, we consider dephasing effect of stray magnetic fields and atom shot noise effect on the AMO process. The dephasing strength is set as uniformly distributed random numbers $\delta B_z \in [-0.1,0.1]$ mG and $q$ changes according to $q = q_{_{MW}} + (B_z+\delta B_z)^2 \times 277$ Hz/G$^2$ with $q_{_{MW}}$ denoting the level shift induced by a driving microwave field~\cite{Zhao2014Dynamics, Gerbier2006Resonant, Leslie2009Amplification}. We assume that the initially prepared atom number fluctuation of the condensate is uniformly distributed in the range $[N-\sqrt{N}, N+\sqrt{N}]$. The numerical simulation results are presented by the black solid line in Fig.~\ref{fig:4}(a). Clearly, the generalized squeezing parameter monotonically decreases to a lowest value close to $2/N$, after the three multilevel oscillations. The deviation of the minimal $\xi^2$ from the singlet state's value of zero is due to the odd atom numbers in the condensate, whose lowest value $\xi^2=2/N$ ($N$ is an odd integer) for its ground state $|l=1,m=0\rangle$. Allowed to distinguish the odd and even number of atoms by postselection~\cite{note4}, we find the even number condensate continues decreasing to a value much smaller than $1/N$ and very close to the ideal case (grey dashed line), indicating the formation of the many-body singlet state {with a very high fidelity above 99\%} and the robustness of the AMO process.

Second, we consider the relaxation and the dephasing effects of stray magnetic fields on the AMO process. Without loss of generality, we consider the external transversal stray magnetic field is just along the $x$-axis. The effective Hamiltonian becomes
\begin{eqnarray}\label{eq:relax}
H_e=c'_2 \frac{\bm{L}^2}{N} -q a_0^{\dagger}a_0 - p L_z -h L_x,
\end{eqnarray}
where $p=-\gamma (B_z+\delta B_z)$ with a moderate bias $B_z=0.85$ G and the gyromagnetic ratio $\gamma = - 0.7$ MHz/G is the linear Zeeman splitting, and $h=-\gamma \delta B_x$ is for the transversal magnetic field. {We assume that $\delta B_x$ is also uniformly distributed in $[-0.1,0.1]$ mG.} We are limited by the computational power to $N=100$, due to the explosion of the Hilbert space introduced by the $L_x$. The numerical results are shown in Fig.~\ref{fig:4}(b). We find that the dynamics of the parameter $\xi^2$ (with negligible error bars) overlaps with the ideal one, demonstrating the stay magnetic fields within $0.1$ mG rarely affect the multilevel oscillations. In fact, the final fidelity to the singlet state is still higher than $99\%$.

Finally, we take the atom loss and dephasing effects into consideration. The dynamics must be depicted by the following master equation,
\begin{eqnarray}
\label{eq:5}
\dot{\rho}&=&-i[H_e,\rho] +\Gamma\sum_{m}(2a_m \rho a_m^\dagger - a_m^\dagger a_m \rho -\rho a_m^\dagger a_m), \quad
\end{eqnarray}
where we take $\Gamma=0.005$ s$^{-1}$ and $H_e$ is the Hamiltonian in Eq.~(\ref{eq:1}). In the giant Hilbert space spanned by $|N,l,m\rangle$, we carry out numerical simulations for an ``initial" atom number $N(t=0.9 \; \text{s} )=100$, focusing on the multilevel oscillation process. As shown in Fig.~\ref{fig:4}(b), the generalized spin squeezing parameter $\xi^2$ (with negligible error bars) also reaches $1/N$ but higher than the ideal case. Here we note that the final fidelity to the singlet state drops down to $70\%$ due to the atom loss, but it can be easily remedied by a postselection procedure and the fidelity is improved to a value higher than 99\%~\cite{note2}.

{In conclusion, we almost perfectly generate the long-sought massively entangled states, both for the many-body singlet and twin-Fock states in an antiferromagnetic $^{23}$Na spin-1 condensate with the AMO and AMOA processes.} The numerical simulations show that the generation efficiencies of both states are over $96\%$ in a few seconds and in moderate magnetic fields (from milliGauss to Gauss). Under realistic experimental conditions, the AMO process is robust against the dephasing and relaxation noises of stray magnetic fields, the atom shot noise, and the atom loss. {It is worthy to explore in the future the potential of the multilevel oscillations, to replace the adiabatic evolution, near a quantum critical point in many physical systems, e.g., ferromagnetic $^{87}$Rb spin-1 condensates, two-level quantum systems, and adiabatic spin quantum computers~\cite{Luo2017Deterministic, Peng2008Quantum}}. The generated many-body spin singlet state provides a stepping stone to reach the Heisenberg limit gradient magnetometer~\cite{Toth2013Macroscopic} and the twin-Fock state can be directly utilized to measure the external magnetic field beyond the standard quantum limit~\cite{Duan2013Generation, Luo2017Deterministic}.

\begin{acknowledgments}
We thank R.Q. Wang and L. You for inspiring discussions. This work was supported by the NSFC (Grant No. 91836101, No. 11574239, No. 11434011, and No. 11674334) and by the Open Research Fund Program of the State Key Laboratory of Low Dimensional Quantum Physics under Grant No. KF201614.
\end{acknowledgments}



\begin{thebibliography}{64}
\bibitem[{not({\natexlab{b}})}]{note4}
    \bibinfo{note}{In each experiment, one may measure the generalized spin
      squeezing parameter $\xi^2$ at the end of the multilevel oscillations. If
      $\xi^2<2/N$, the number of atoms must be even; otherwise the number of atoms
      is odd, because the minimum of the generalized spin squeezing parameter
      equals $2/N$ (for odd $N$).}
\bibitem[{not({\natexlab{c}})}]{note2}
    \bibinfo{note}{Note that we do not consider the atom loss in the adiabatic
      process, which makes the final squeezing parameter and the fidelity a little
      different, but does not affect the final fidelity after the postselection.}
\end{thebibliography}

\end{document}